\documentstyle[12pt]{article}
\newcommand{\extraspace}{\addtolength{\abovedisplayskip}{2mm}
\addtolength{\belowdisplayskip}{2mm}
\addtolength{\abovedisplayshortskip}{2mm}
\addtolength{\belowdisplayshortskip}{2mm}}
\newcommand{\be}{\begin{equation}\extraspace}

\newcommand{\ee}{\end{equation}}

\newcommand{\bea}{\begin{eqnarray}\extraspace}
\newcommand{\beastar}{\begin{eqnarray*}\extraspace}
\newcommand{\eea}{\end{eqnarray}}
\newcommand{\eeastar}{\end{eqnarray*}}

\newcommand{\eps}{\epsilon}

\newcommand{\up}{\uparrow}
\newcommand{\down}{\downarrow}

\def\lefthook{{\vrule height5pt width0.4pt depth0pt}}
\def\righthook{{\vrule height5pt width0.4pt depth0pt}}
\def\leftrighthookfill{$\mathsurround=0pt \mathord\lefthook
\hrulefill\mathord\righthook$}
\def\underhook#1{\vtop{\ialign{##\crcr$\hfil\displaystyle{#1}\hfil$\crcr
\noalign{\kern-1pt\nointerlineskip\vskip2pt}
\leftrighthookfill\crcr}}}

\newcommand{\newsection}[1]{
\vspace{12mm}
\pagebreak[3]
\addtocounter{section}{1}
\setcounter{equation}{0}
\setcounter{subsection}{0}
\setcounter{footnote}{0}
\begin{flushleft}
{\large\bf \thesection. #1}
\end{flushleft}
\nopagebreak}

\def\ZZ{Z\!\!\!Z} 		
\def\eql{~=~}

\def\bK{{\bf K}} \def\bQ{{\bf Q}}
\def\wt#1{{\widetilde{#1}}}
\def\scr#1{{\scriptstyle{#1}}}
\def\txt#1{{\textstyle{#1}}}

\expandafter\ifx\csname amssym.def\endcsname\relax \else\endinput\fi
\expandafter\edef\csname amssym.def\endcsname{%
       \catcode`\noexpand\@=\the\catcode`\@\space}
\catcode`\@=11
\def\undefine#1{\let#1\undefined}
\def\newsymbol#1#2#3#4#5{\let\next@\relax
 \ifnum#2=\@ne\let\next@\msafam@\else
 \ifnum#2=\tw@\let\next@\msbfam@\fi\fi
 \mathchardef#1="#3\next@#4#5}
\def\mathhexbox@#1#2#3{\relax
 \ifmmode\mathpalette{}{\m@th\mathchar"#1#2#3}%
 \else\leavevmode\hbox{$\m@th\mathchar"#1#2#3$}\fi}
\def\hexnumber@#1{\ifcase#1 0\or 1\or 2\or 3\or 4\or 5\or 6\or 7\or 8\or
 9\or A\or B\or C\or D\or E\or F\fi}

\font\tenmsa=msam10 scaled \magstep1
\font\sevenmsa=msam7 scaled \magstep1
\font\fivemsa=msam5 scaled \magstep1
\newfam\msafam
\textfont\msafam=\tenmsa
\scriptfont\msafam=\sevenmsa
\scriptscriptfont\msafam=\fivemsa
\edef\msafam@{\hexnumber@\msafam}
\mathchardef\dabar@"0\msafam@39
\def\dashrightarrow{\mathrel{\dabar@\dabar@\mathchar"0\msafam@4B}}
\def\dashleftarrow{\mathrel{\mathchar"0\msafam@4C\dabar@\dabar@}}

\def\ulcorner{\delimiter"4\msafam@70\msafam@70 }
\def\urcorner{\delimiter"5\msafam@71\msafam@71 }
\def\llcorner{\delimiter"4\msafam@78\msafam@78 }
\def\lrcorner{\delimiter"5\msafam@79\msafam@79 }
\def\yen{{\mathhexbox@\msafam@55}}
\def\checkmark{{\mathhexbox@\msafam@58}}
\def\circledR{{\mathhexbox@\msafam@72}}
\def\maltese{{\mathhexbox@\msafam@7A}}

\font\tenmsb=msbm10 scaled \magstep1
\font\sevenmsb=msbm7 scaled \magstep1
\font\fivemsb=msbm5 scaled \magstep1

\newfam\msbfam
\textfont\msbfam=\tenmsb
\scriptfont\msbfam=\sevenmsb
\scriptscriptfont\msbfam=\fivemsb
\edef\msbfam@{\hexnumber@\msbfam}
\def\Bbb#1{{\fam\msbfam\relax#1}}
\def\widehat#1{\setbox\z@\hbox{$\m@th#1$}%
 \ifdim\wd\z@>\tw@ em\mathaccent"0\msbfam@5B{#1}%
 \else\mathaccent"0362{#1}\fi}
\def\widetilde#1{\setbox\z@\hbox{$\m@th#1$}%
 \ifdim\wd\z@>\tw@ em\mathaccent"0\msbfam@5D{#1}%
 \else\mathaccent"0365{#1}\fi}
\font\teneufm=eufm10 scaled \magstep1
\font\seveneufm=eufm7 scaled \magstep1
\font\fiveeufm=eufm5 scaled \magstep1
\newfam\eufmfam
\textfont\eufmfam=\teneufm
\scriptfont\eufmfam=\seveneufm
\scriptscriptfont\eufmfam=\fiveeufm

\font\teneufb = eufb10 scaled \magstep1

\newfam\eufbfam
\textfont\eufbfam=\teneufb

\font\tencmbsy  = cmbsy10 scaled \magstep1

\font\sevencmbsy = cmbsy7 scaled \magstep1
\font\fivecmbsy = cmbsy5 scaled \magstep2

\newfam\cmbsyfam
\textfont\cmbsyfam=\tencmbsy
\scriptfont\cmbsyfam=\sevencmbsy
\scriptscriptfont\cmbsyfam=\fivecmbsy

\csname amssym.def\endcsname

\def\ZZ{{\Bbb Z}}

\def\Journal#1#2#3#4{{#1} {{#2}} ({#3}) {#4}}
\def\AdP{Adv.\ Phys.\ }

\def\NPB{Nucl.\ Phys.\ B}
\def\PRL{Phys.\ Rev.\ Lett.\ }
\def\PRB{Phys.\ Rev.\ B}

\def\MPLB{Mod.\ Phys.\ Lett.\ B}

\begin{document}
\baselineskip=17pt

\vskip 1.5cm
\begin{center}

{\Large $K$-matrices for non-abelian quantum Hall states}
\vskip 2.5cm
{\large Eddy Ardonne$^1$, Peter Bouwknegt$^2$, Sathya Guruswamy$^1$}
\vskip .2cm
{\large and Kareljan Schoutens$^1$}
\vskip .3cm
{\sl ${}^1$ Institute for Theoretical Physics \\
     University of Amsterdam \\
     Valckenierstraat 65 \\
     1018 XE  Amsterdam, THE NETHERLANDS}
\vskip .3cm
{\sl ${}^2$ Department of Physics and Mathematical Physics \\
     University of Adelaide \\
     Adelaide, SA 5005, AUSTRALIA}

\vskip 1cm
{\bf Abstract}
\end{center}
Two fundamental aspects of so-called non-abelian quantum
Hall states (the $q$-pfaffian states and more general)
are a (generalized) pairing of the participating electrons
and the non-abelian statistics of the quasi-hole excitations.
In this paper, we show that these two aspects are linked 
by a duality relation, which can be made manifest by
considering the $K$-matrices that describe the exclusion 
statistics of the fundamental excitations in these systems. 

\baselineskip=15pt
{\small

}
\vfill
\leftline{ITFA-99-18}
\leftline{ADP-99-30/M85}
\noindent {{\tt cond-mat/9908285} \hfill  August 1999}

\newpage

\baselineskip=17pt

\newsection{Introduction}

In the description of low-energy excitations over abelian 
fractional quantum Hall (fqH) states, an important role is played
by the so-called $K$-matrices, which characterize the topological 
order of the fqH state (see \cite{We} for a review). 
These $K$-matrices act as 
parameters in Landau-Ginzburg-Chern-Simons (LGCS) field theories 
for bulk excitations and in the chiral Conformal Field Theories
(CFT) that describe excitations at the edge. At the same time, the
$K$-matrices provide the parameters for the fractional exclusion 
statistics (in the sense of Haldane \cite{Hal}) of the edge excitations over
the fqH states. For a simple example 
of this, consider the Laughlin fqH state at filling fraction 
$\nu={1 \over m}$, with $K$-matrix equal to the number $m$. Following
the analysis in \cite{ES}, one finds that the parameters $g_e$ and
$g_{\phi}$ that characterize the exclusion statistics of the edge 
electrons (of charge $Q=-e$) and edge quasi-holes ($Q=+{e \over m}$), 
respectively, are given by
\be
g_e \eql \bK \eql  m \ ,  \qquad\qquad 
g_{\phi}\eql \bK^{-1} \eql {1 \over m} \ .
\label{Klaugh}
\ee
In this paper, we shall denote these and similar parameters by 
$\bK_e$ and $\bK_{\phi}$, respectively.

The $(e,\phi)$ basis for edge excitations is natural in view
of the following statements about duality and completeness.
The {\it particle-hole duality}\ between the edge electron
and quasi-hole excitations is expressed through the relation 
$\bK_{\phi}=\bK_e^{-1}$ and through the absense of mutual exclusion 
statistics between the two. It leads to the following 
relation between the 1-particle distribution functions
$n_e(\eps)$ and $n_\phi(\eps)$
\be
m \, n_e(\eps) \eql 1 - {1 \over m}\, n_{\phi} (-{\eps \over m})\ .
\label{eqPBb}
\ee
In physical terms, the duality implies that the absence of 
edge electrons with energies $\eps<0$ is equivalent to the 
presence of edge quasi-holes with positive energies, and vice 
versa.
By the {\it completeness}\ of the $(e,\phi)$ basis we mean that the 
collection of all multi-$e$, multi-$\phi$ states span a basis of 
the chiral Hilbert space of all edge excitations. In mathematical
terms, this is expressed by a formula that expresses the partition 
sum for the edge excitations as a so-called Universal Chiral 
Partition Function (UCPF) (see, e.g., \cite{BM} and
references therein) based on the matrix 
\be
\bK_e \oplus \bK_\phi = \left(\begin{array}{cc} m & 0 \\ 
                               0     & {1 \over m} 
\end{array} \right) \ .
\ee
We refer to \cite{ES} for an extensive discussion of these results,
and to \cite{FK} for an extension to (abelian) fqH states with
more general $n\times n$ $K$-matrices.

Turning our attention to non-abelian quantum Hall states, we 
observe that the chiral CFT for the edge excitations are not 
free boson theories. This implies that for these states the 
notion of a $K$-matrix needs to be generalized. In this paper, 
we shall show that the exclusion statistics
of the fundamental edge `quasi-hole' and `electron' excitations
over non-abelian quantum Hall states give rise to matrices 
that are closely analogous to the $K$-matrices in the abelian case,
and we therefore refer to these matrices as the $K$-matrices for 
non-abelian quantum Hall states. We 
present results for the $q$-pfaffian state \cite{MR}, for the 
so-called parafermionic quantum Hall states \cite{RR} and for 
the non-abelian spin-singlet states recently proposed in \cite{AS}.
A more detailed account, including explicit derivations of the
results presented here, will be given elsewhere \cite{ABS}.

\newsection{Pfaffian quantum Hall states for spinless electrons}

The so-called $q$-pfaffian quantum Hall states, at filling fraction
$\nu={1 \over q}$ were proposed in 1991 by Moore and Read \cite{MR} and 
have been studied in considerable detail \cite{GWW,NW,RG}. 
The charged spectrum contains fundamental 
quasi-holes of charge ${e \over 2q}$ and electron-type excitations with
charge $-e$ and fermionic braid statistics. The edge CFT can be written 
in terms of a single chiral boson and real (Majorana) fermion, leading
to a central charge $c={3 \over 2}$. The exclusion statistics of the
edge quasi-holes were studied in \cite{Sc,GS}. In the second reference, 
it was found that the thermodynamics of the edge quasi-hole can be
described by the following equations
\be
{\lambda_0-1 \over \lambda_0}  \lambda_0 \lambda^{-{1 \over 2}} \eql 1 \ ,
\qquad\quad
{\lambda -1 \over \lambda} 
   \lambda_0^{-{1\over 2}} \lambda^{q+1 \over 4q} \eql x \ ,
\label{IOWpf}
\ee
with $x=e^{\beta(\mu_{\phi}-\eps)}$. Comparing this with the 
general form of the Isakov-Ouvry-Wu (IOW) equations for particles 
with exclusion statistics matrix $\bK$ \cite{IOW},
\be
\left( {\lambda_a-1 \over \lambda_a} \right) 
\prod_b \lambda_b^{K_{ab}} \eql x_a \ ,
\label{IOW}
\ee
one identifies 
\be
\bK_{\phi} \eql 
\left(\begin{array}{cc}  \hphantom{-}1 & -\scr{1 \over 2} \\
                        -\scr{1 \over 2} & \hphantom{-}{q+1 \over 4q}
\end{array} \right)
\label{eqPBa}
\ee
as the statistics matrix for particles $(\phi_0,\phi)$, where $\phi$ is 
the edge quasi-hole of charge ${e \over 2q}$. The other particle 
$\phi_0$ does not carry any charge or energy and is called a 
{\it pseudo-particle}. The presence of this particle accounts for the 
non-abelian statistics of the physical particle $\phi$ \cite{GS,BCR}. 
Eliminating $\lambda_0$ from equations (\ref{IOWpf}) gives 
\be
(\lambda-1)(\lambda^{1 \over 2}-1) \eql x^2 \lambda^{3q-1 \over 2q} \ ,
\ee
in agreement with \cite{Sc}.

The duality between the $\phi$ and $e$ excitations over the
$q$-pfaffian state was first discussed in \cite{Sc}, where it was also
shown how the correct spectrum of the edge CFT is
reconstructed using the $e$ and $\phi$ quanta. Here we present a
discussion at the level of $K$-matrices where, as in
eqn.\ (\ref{Klaugh}), the dual sector is reached by inverting the matrix
$\bK_\phi$ of eqn.\ (\ref{eqPBa}).

Starting from the IOW equations (\ref{IOW}), and 
denoting by $\lambda^\prime_a$ and $x_a^\prime$ the quantities 
corresponding to $\bK^\prime=\bK^{-1}$, we have the 
correspondence \cite{ABS}
\be
\lambda_a^\prime \eql {\lambda_a \over \lambda_a-1} \ , 
\qquad\quad
x_a^\prime \eql \prod_b x_b^{-K_{ab}^{-1}}\ .
\ee
which leads, among others, directly to the (appropriate generalization of)
eqn.\ (\ref{eqPBb}).

For the $q$-pfaffian state, we define the $K$-matrix for the 
electron sector ($e$-sector) to be the inverse of $\bK_{\phi}$
\be 
\bK_e \eql \bK_{\phi}^{-1} \eql 
\left(\begin{array}{cc} q+1 & 2q \\
                        2q & 4q
\end{array} \right) \ .
\label{eqPBaa}
\ee
Inspecting the right hand sides of the duality-transformed
IOW equations, we find $x^\prime_1=y$ and $x^\prime_2=y^2$,
with $y=e^{\beta(\mu_e-\eps')}$, where $\eps'=-2q\eps$ and
$\mu_e=-2q \mu_{\phi}$ (i.e., $y=x^{-2q}$),
indicating that the two particles in the $e$-sector carry 
charge $Q=-e$ and $Q=-2e$, respectively. We shall denote these
particles by $\Psi_1$ and $\Psi_2$. The first particle is 
quickly identified with the edge electron, with self-exclusion 
parameter equal to ${q+1}$. The presence of a `composite'
particle $\Psi_2$ of charge $-2e$ has its origin in the fundamental 
electron pairing that is implied by the form of the pfaffian wave
function. 

As before, there is no mutual exclusion statistics between the
$e$- and the $\phi$-sectors. In ref.~\cite{ABS}, we show that
the conformal characters of the edge CFT can be cast in the UCPF 
form with matrix $\bK_e \oplus \bK_\phi$. The value $c={3 \over 2}$
of the central charge follows as a direct consequence \cite{BCR}. 

We remark that the usual relations between the charge vector 
$\bQ^T_e=(-e,-2e)$, $\bQ^T_{\phi}=(0,{e \over 2q})$, the matrices $\bK_e$,
$\bK_{\phi}$ and the filling fraction $\nu$ are satisfied in this 
non-abelian context
$$
\bK_e \eql \bK_\phi^{-1} \,,\qquad \bQ_e \eql - \bK_\phi^{-1}\cdot \bQ_\phi\ ,
$$
\be
\nu e^2 \eql \bQ^T_\phi \cdot \bK^{-1}_\phi \cdot \bQ_\phi \eql
 \bQ^T_e \cdot \bK^{-1}_e \cdot \bQ_e  \ .
\ee
These relations, which hold in all examples discussed in this paper,
together with the UCPF form of the conformal 
characters, motivate our claim that $\bK_e$ and $\bK_\phi$ are
the appropriate generalizations of the $K$-matrices in this
non-abelian setting.
  
We shall now proceed and link the composite $Q=-2e$
particle in the electron sector to the supercurrent that is familiar 
in the context of the BCS theory for superconductivity. 
For this argument, it is useful to follow the $q$-pfaffian state as 
a function of $q$ with $0\leq q \leq 2$. This procedure can be 
interpreted in terms of a flux-attachment transformation \cite{GWW,RG}.
For $q=2$ we have the fermionic pfaffian state at $\nu={1 \over 2}$ 
and $q=1$ gives a bosonic pfaffian state at $\nu=1$.
In the non-magnetic limit, $q\to 0$, we recognize the pfaffian 
wave function as the coordinate space representation of a specific 
superconducting BCS state with complex $p$-wave pairing \cite{GWW,RG}. 
In the limit $q\to 0$, the particle $\Psi_2$ has exclusion parameter 
$g=0$ and can, as we shall argue, be associated with the supercurrent
of the superconducting state.  [In fact, the mutual exclusion statistics
between $\Psi_2$ and $\Psi_1$ vanishes as well in the limit $q\to 0$.]
The exclusion statistics parameter for the $\phi$-particle diverges for 
$q\to 0$, meaning that in this limit the $\phi$-sector no longer 
contributes to the physical edge spectrum.   
 
In a quantum state that has all electrons paired, the fundamental 
flux quantum is ${h \over 2e}$. Piercing the sample with precisely this
amount of flux leads to a quasi-hole excitation of charge 
${e \over 2q}$. This excitation also contains a factor which
acts as the spin-field with respect to the neutral fermion in the 
electron-sector; the latter factor is at the origin of the non-abelian
statistics \cite{MR,NW}. For $q=1,2, \ldots$ the quasi-hole charge 
${e \over 2q}$ is
the lowest charge that we consider: the excitations of charge $-e$ 
and $-2e$ correspond to flux insertions that are a (negative) integer 
multiple of the flux quantum. However, for $q\ll 1$, the quasi-hole 
charge is larger than $e$, $2e$, and we conclude that the 
fundamental excitations in the $e$-sector correspond to an
insertion of a fraction of the flux-quantum, in other words, to
a situation where the boundary conditions for the original 
`electrons' have been twisted.

For definiteness, let us put $q={1 \over N} $ with $N$ a large integer.
A $\Psi_2$ quantum state of charge $-2e$ 
then corresponds to a flux insertion  
of $-{2qh \over e}=-{2h \over Ne}$. In the absence of any $\Psi_1$ quanta, 
this quantum state for the $\Psi_2$ particle can be filled up to a maximum
of $n^{\rm max}={1 \over 4q}={N \over 4}$ times. [This follows
from the self-exclusion parameter $g=4q$ and the result that in Haldane's
statistics, $n^{\rm max}={1 \over g}$.] The amount of flux that corresponds 
to this maximal occupation equals $-{2h \over Ne} {N \over 4}
= -{h \over 2e}$, which is precisely the flux quantum.

Summarizing, we see that the insertion of a single flux quantum 
${h \over 2e}$ gives rise to a quasi-hole ($\phi$) excitation, while 
(negative) fractions of the flux quantum (between $-{h \over 2e}$ and 0) 
give rise multiple occupation of the $\Psi_2$ modes.

In the description of a BCS superconductor, the excitation that is
induced by the insertion of a fraction of the flux quantum ${h \over 2e}$,
is precisely the supercurrent that screens the imposed 
flux. Comparing the two pictures, we see that in the limit $q\to 0$, the 
$\Psi_2$ excitations in the $q$-pfaffian state reduce to supercurrent 
excitations in the limiting BCS superconductor. 
In an earlier study \cite{GWW}, the neutral fermionic excitation over
the $q$-pfaffian (which is not elementary in our $(e,\phi)$ description),
in the limit $q\to 0$, has been identified with the pair breaking
excitation of the superconducting state.

A similar reasoning applies to the Laughlin state at $\nu={1\over
m}$, where it links the charge $-e$ excitations at finite $m$ to the
supercurrent of the limiting superfluid boson state at $m=0$.

In a somewhat different physical picture, the edge particle $\Psi_2$
is the one that is excited in the process of Andreev reflection off the
edge of a sample in the $q$-pfaffian state. It will be interesting to
explore in some detail such processes for the $q=2$ pfaffian state.

\newsection{Pseudo-spin triplet pfaffian quantum Hall states}

As a generalization of the results in the previous section,
we consider a $q$-pfaffian state for particles with an internal
(double-layer or pseudo-spin) degree of freedom. 
The wave function of this state is the
Halperin two-layer state with labels $(q+1,q+1,q-1)$. This state has 
$K$-matrices 
\be 
\bK_e \eql  
\left(\begin{array}{cc} q+1 & q-1 \\
                        q-1 & q+1
\end{array} \right) \ ,
\qquad
\bK_\phi \eql  {1 \over 4q} 
\left(\begin{array}{cc} \hphantom{-}q+1 & -q+1 \\
                        -q+1 & \hphantom{-}q+1
\end{array} \right) \ ,
\ee
describing excitations $(\Psi_\up,\Psi_\down)$ of charge
$-e$ and $(\phi_\up,\phi_\down)$ of charge ${e \over 2q}$,
respectively. There are no pseudo-particles and these states 
are abelian quantum Hall states.

We shall now argue that, based on the analogy with the $q$-pfaffian 
states for spinless electrons, for these states we can identify 
supercurrent type excitations in the $e$-sector and, by duality, 
reformulate the $\phi$-sector in terms of one physical quasi-hole 
and two pseudo-particles. We stress that this reformulation does
not change the physical interpretation; in particular, although the new
formulation employs two pseudo-particles, it still refers to an
abelian quantum Hall state.

In the limit $q\to 0$, the $(q+1,q+1,q-1)$ paired wave function 
reduces to a form that
can be interpreted as a (complex) $p$-wave pseudo-spin triplet state
\cite{RG}. Taking $q={1 \over N}$ as before, we can look for excitations
in the $e$-sector that describe the response to the insertion of
fractional flux, and that will smoothly connect to the supercurrent 
at $q=0$. Inspecting the matrix $\bK_e$, we see that the two quanta
$\Psi_\up$ and $\Psi_\down$ each have self-exclusion parameter
approaching $g=1$, and can by themselves not screen more than 
an amount of flux equal to ${h \over Ne}$. However, due to the strong 
negative mutual exclusion statistics, an excitation that is effectively a
pair of one $\Psi_\up$ and one $\Psi_\down$ particle can screen
a much larger amount of flux.

To formalize this consideration, we introduce a particle $\Psi_3$,
defined as a pair $(\Psi_\up\Psi_\down)$. Following a general
construction presented in \cite{ABS}, we derive a new $K$-matrix
for the extended system $(\Psi_\up,\Psi_\down,\Psi_3)$
\be 
\wt{\bK}_e \eql 
\left(\begin{array}{ccc} q+1 & q   & 2q \\
                         q   & q+1 & 2q \\
                         2q  & 2q  & 4q  
\end{array} \right) \ .
\label{eqPBf}
\ee
[This choice of $K$-matrix guarantees an equivalence between the
$(\Psi_\up, \Psi_\down)$ and the $(\Psi_\up,\Psi_\down,\Psi_3)$
formulations.] On the basis of the extended matrix $\wt{\bK}_e$, we 
identify the
supercurrent excitations as before: a single $\Psi_3$ quantum
requires flux ${2h \over Ne}$, and with a maximal filling of
$n^{\rm max}={1 \over 4q}={N \over 4}$, we see that the 
$\Psi_3$ quanta can `absorb' an amount of flux equal to 
${h \over 2e}$. In the limit $q\to 0$, the $\Psi_3$ quanta
are identified with the supercurrent quanta, which have the ability 
to screen a full quantum ${h \over 2e}$ of applied flux.
Inverting $\wt{\bK}_e$, we obtain
\be
\wt{\bK}_{\phi} \eql
\left(\begin{array}{ccc} \hphantom{-}1 &  \hphantom{-}0& -\scr{1 \over 2} \\
                         \hphantom{-}0 &  \hphantom{-}1& -\scr{1 \over 2} \\
   -\scr{1 \over 2}  & -\scr{1 \over 2}  & \hphantom{-}{2q+1 \over 4q}
\end{array} \right) \ ,
\label{eqPBg}
\ee
with associated parameters
\be
x_1 \eql \left( {y_\up \over y_\down} \right)^{1 \over 2} \ , \qquad
x_2 \eql \left( {y_\down \over y_\up} \right)^{1 \over 2} \ , \qquad
x_3 \eql (y_\up y_\down)^{-{1 \over 4q}} \ ,
\ee
where $y_{\up,\down} = \exp( \beta(\mu_{\up,\down} - \epsilon) )$.
The fact that $x_{1,2}$ do not depend on the energy parameter
$\eps$ makes clear that these are pseudo-particles. As such they 
account for degeneracies in states that contain more than one 
$\phi_3$-quantum. Despite this appearance, by 
construction it is clear that the braid statistics of these degenerate
excitations are abelian.

At the level of the edge CFT, the two different formulations of
the $(q+1,q+1,q-1)$ theory are easily understood. In the usual 
abelian formulation, the edge CFT is written in terms of a (charge)
boson $\varphi_c$ plus a Dirac fermion, whose (dimension-${1 \over 8}$) 
spin field has abelian statistics. 
The alternative formulation employs $\varphi_c$
plus two real fermions (called $\psi_e$ and $\psi_o$ in \cite{RG}).
The two pseudo-particles that we obtained describe the non-abelian
statistics of the spin-fields of the real (Ising) fermions 
$\psi_e$ and $\psi_o$ separately. The actual chiral Hilbert space
of the edge CFT is however a subspace of the Hilbert space of
the (Ising)$^2$ CFT, and the braid statistics in this subspace are 
all abelian. 

The various phase transitions described in \cite{RG} are easily
traced in the statistics matrices. The transition of the $o$ spins
into the strong-pairing phase decouples one row of the matrices
(\ref{eqPBf}) and (\ref{eqPBg}), turning them into the matrices
(\ref{eqPBaa}), (\ref{eqPBa}) corresponding to the $q$-pfaffian state.
A subsequent transition of the $e$ spins into the strong-pairing 
phase leaves only the edge excitations $\Psi_3$ and further reduces 
the matrix to $\bK_e=4q$, appropriate for a Laughlin state of charge 
$-2e$ particles at filling $\nu={1 \over q}$.

\newsection{Parafermionic quantum Hall states: generalized pairing}

In \cite{RR}, Read and Rezayi proposed a series of non-abelian 
quantum Hall states based on order-$k$ clustering of spinless electrons.
The wave functions for these states are constructed with help of
the well-known $\ZZ_k$ parafermions. The general state of \cite{RR},
labeled as $(k,M)$, has filling fraction $\nu={k \over kM+2}={1 \over q}$
with $q=M+{2 \over k}$. Fermionic quantum Hall states are obtained for 
$M$ an odd integer. [For $M=0$ we have a bosonic state with 
$SU(2)_k$ symmetry.]
For $k=1,2$ these new states reduce to the Laughlin
$(m=M+2)$ and $q$-pfaffian ($q=M+1$) states, respectively.

In \cite{GS}, the matrices $\bK_{\phi}$ were identified for general $(k,M)$.
Here we illustrate the $K$-matrix structure for $k=3$, where we have
\be
\bK_{\phi} \eql
\left(\begin{array}{ccc} 
  \hphantom{-}1         &   -\scr{1 \over 2}     &   \hphantom{-}0  \\
  -\scr{1 \over 2}  &       \hphantom{-}1       & -\scr{1 \over 2} \\
  \hphantom{-}0         & -\scr{1 \over 2}& \hphantom{-}{3q + 1 \over 9q}
\end{array} \right)\,,\qquad
\bQ^T_\phi \eql (0,0,\txt{e\over 3q}) \ ,
\ee
for two pseudo-particles and a physical quasi-hole of
charge ${e \over 3q}$. Inverting this matrix gives
\be 
\bK_e \eql 
\left(\begin{array}{ccc} \strut q+\scr{2 \over 3} & 2q+\scr{2 \over 3}  & 3q \\
                  \strut 2q+\scr{2 \over 3}  & 4q+\scr{4 \over 3} & 6q \\
                         3q  & 6q  & 9q  
\end{array} \right) \,,\qquad
\bQ^T_e \eql (-e,-2e,-3e)\ .
\ee
Again putting $q={1 \over N}$, with $N$ large, we can repeat the previous
arguments. Clearly, the $\Psi_3$ quanta of charge $-3e$ act as the 
`supercurrent' for the 3-electron clustering. One such quantum 
requires a flux of $-{3h \over eN}$, and with $n^{\rm max}={N \over 9}$, 
the total flux that can be absorbed equals $-{h \over 3e}$ as expected.

We remark that, for $q\to0$,
the excitations $\Psi_{1,2}$ have fractional exclusion
statistics parameters, in agreement with the fact that the $k=3$ state
at $q=0$ has $M=-{2 \over 3}$ and is thus not fermionic. What one has
instead is an `anyonic' superconductor with Cooper clusters of charge
$-3e$ and cluster-breaking excitations with fractional exclusion
statistics.

\newsection{Non-abelian spin-singlet quantum Hall states}

In \cite{AS}, two of the present authors introduced a series
of non-abelian spin singlet (NASS) states. The states are labeled
as $(k,M)$ and have filling fraction $\nu={2k \over 2kM+3}
={1 \over q}$ with $q=M+{3 \over 2k}$. The wave functions,
which are constructed as conformal blocks of higher rank
(Gepner) parafermions, have a BCS type factorized form, where
the factors describe a $k$-fold spin-polarized clustering of 
electrons of given spin and the formation of a spin-singlet
with $2k$ participating electrons. [See \cite{AS} for an 
example and \cite{ABS} for general and explicit expressions
for the wave functions.]

For $k=1$ the spin-singlet states are abelian with $K$-matrices
given by
\be 
\bK_e \eql  
\left(\begin{array}{cc} q+\scr{1\over 2} & q-\scr{1\over 2} \\
                        q-\scr{1\over 2} & q+\scr{1\over 2}
\end{array} \right) \ ,
\qquad
\bK_\phi \eql  {1 \over 2q} 
\left(\begin{array}{cc} \hphantom{-}q+\scr{1\over 2} & -q+\scr{1\over 2} \\
                        -q+\scr{1\over 2} & \hphantom{-}q+\scr{1\over 2}
\end{array} \right) \ .
\ee
We remark that, as for the Laughlin
series, there is a self-duality in the sense that 
$\bK_{e}(M) = \bK_{\phi}(M')$ with 
\be
M \eql - {3M'+4 \over 2M'+3}\,, \qquad \qquad 
\left( q \eql {1\over4q'} \right)
\ee
One of the self-dual points is $M=M'=-1$ ($q=q'={1\over2}$),
corresponding to two decoupled $\nu=1$ systems for spin
up and down.

In a forthcoming paper \cite{ABS} we present a detailed
derivation of the $K$-matrix structure for the general
NASS states, where we obtain `minimal' $K$-matrices of size 
$2k \times 2k$. The 
matrix $\bK_e$ describes fully polarized composites (of both
spins) of $1,2,\ldots,k$ quasi-electrons, while the
matrix $\bK_{\phi}$ describes a spin-doublet of physical,
fractionally charged quasi-holes ($Q={e \over 4q}$) and 
a collection of $2(k-1)$ pseudo-particles that take care 
of the non-abelian statistics. The simplest non-trivial 
example is the result for $k=2$, ($M=q-{3 \over 4}$)
\bea
&&
\bK_{\phi} \eql 
\left(\begin{array}{cccc} 
\hphantom{-}\scr{4\over3} & \hphantom{-}\scr{2\over3} & 
-\scr{2\over3} & -\scr{1\over3} \\
\hphantom{-}\scr{2\over3} & \hphantom{-}\scr{4\over3} & 
-\scr{1\over3} & -\scr{2\over3} \\
-\scr{2\over3} & -\scr{1\over3} & \hphantom{-}{28q+3\over 48q} & 
  -{4q-3\over48q} \\
-\scr{1\over3} & -\scr{2\over3} & -{4q-3\over48q} & 
  \hphantom{-}{28q+3\over 48q}  
\end{array}\right) \,, \qquad
\bQ^T_\phi \eql (0,0,\txt{e\over 4q},\txt{e\over 4q}) \ ,
\\&&
\bK_e \eql  
\left(\begin{array}{cccc} 
\strut q+\scr{5 \over 4} & q-\scr{3 \over 4} &
2q+ \scr{1\over 2} & 2q - \scr{1\over 2} \\
\strut q-\scr{3 \over 4} & q+\scr{5 \over 4} &
2q- \scr{1\over 2} & 2q + \scr{1\over 2} \\
\strut 2q+ \scr{1\over 2} & 2q - \scr{1\over 2} &
4q+1 & 4q-1 \\
\strut 2q- \scr{1\over 2} & 2q + \scr{1\over 2} &
4q-1 & 4q+1
\end{array} \right) \,, \qquad
\bQ^T_e \eql (-e,-e,-2e,-2e)\ .\nonumber
\eea
In analogy with the reasoning for the pseudo-spin triplet
pfaffian state, one may now consider the composite of the 
$k$-spin-up and the $k$-spin-down components, and determine 
an extended $K_e$-matrix (cf.\ (\ref{eqPBf})). 
For $q\to 0$ one finds that all 
statistical couplings of this composite vanish, and we 
identify it with the supercurrent corresponding to the 
$2k$-electron spin-singlet clustering. The extended $K_e$-matrix 
is invertible and gives a redefined $\phi$-sector with a 
single spinless $\phi$-quantum and $(2k-1)$ pseudo-particles. 
For obtaining a formulation with manifest $SU(2)$-spin symmetry,
a further extension of $\bK_e$ can be considered \cite{ABS}.

\newsection{Conclusions}

In this paper, we have studied the exclusion statistics
of edge excitations over non-abelian quantum Hall states.
From the results we have extracted matrices $\bK_e$ and
$\bK_\phi$ which generalize the well-known $K$-matrices for
abelian quantum Hall states to the non-abelian case.
Note, however, that the torus degeneracy for the non-abelian
case is not given by the abelian result $|{\rm det}(\bK_e)|$, 
but that a further reduction is necessary due to the presence 
of the pseudo-particles. [Compare with \cite{WZ} where such
a reduction was discussed in the context of the parton 
construction of non-abelian quantum Hall states.]

We expect that these 
new $K$-matrices can be used to formulate effective
(edge and bulk) theories for these non-abelian quantum Hall states.
Until now, effective field theories for bulk excitations (of the
LGCS type) have only been obtained for some very special 
cases \cite{FNS}, and it will be most interesting to 
find more systematic constructions.

\bigskip\bigskip
\leftline{\large\bf Acknowledgements}\bigskip

We would like to thank Nick Read for illuminating discussions.
This research was supported in part by the Australian Research Council 
and the foundation FOM of the Netherlands.




\frenchspacing
\baselineskip=16pt
\begin{thebibliography}{11}

\bibitem{We}
X.-G.~Wen, 
\Journal{\AdP}{44}{1995}{405}, [{\tt cond-mat/9506066}].

\bibitem{Hal}
F.D.M.~Haldane, 
\Journal{\PRL}{67}{1991}{937}.

\bibitem{ES}
R.~van Elburg and K.~Schoutens, 
\Journal{\PRB}{58}{1998}{15704}, \newline 
[{\tt cond-mat/9801272}].

\bibitem{BM}
A.~Berkovich and B.~M.~McCoy,
{\it The universal chiral partition function for exclusion statistics},
[{\tt hep-th/9808013}].

\bibitem{FK}
T.~Fukui and N.~Kawakami, 
\Journal{\PRB}{51}{1995}{5239}, [{\tt cond-mat/9408015}].

\bibitem{MR}
G.~Moore and N.~Read,
\Journal{\NPB}{360}{1991}{362}.

\bibitem{RR}
N.~Read and E.~Rezayi, 
\Journal{\PRB}{59}{1999}{8084},
[{\tt cond-mat/9809384}].

\bibitem{AS}
E.~Ardonne and K.~Schoutens,
\Journal{\PRL}{82}{1999}{5096}, \newline 
[{\tt cond-mat/9811352}].

\bibitem{ABS}
E.~Ardonne, P.~Bouwknegt and K.~Schoutens, 
in preparation.

\bibitem{GWW}
M.~Greiter, X.-G.~Wen and F.~Wilczek,
\Journal{\NPB}{374}{1992}{567}.

\bibitem{NW}
C.~Nayak and F.~Wilczek,
\Journal{\PRL}{73}{1994}{2740}.

\bibitem{RG}
N.~Read and D.~Green, 
{\it Paired states of fermions in two dimensions with breaking of 
parity and time-reversal symmetries, and the fractional quantum Hall effect},
[{\tt cond-mat/9906453}].

\bibitem{Sc}
K.~Schoutens, 
\Journal{\PRL}{81}{1998}{1929}, [{\tt cond-mat/9803169}]. 

\bibitem{GS}
S.~Guruswamy and K.~Schoutens,
{\it Non-abelian exclusion statistics}, 
Nucl.\ Phys.\ B, in print, [{\tt cond-mat/9903045}].

\bibitem{IOW}
S.B.~Isakov, \Journal{\MPLB}{8}{1994}{319};  \\
S.~Ouvry, \Journal{\PRL}{72}{1994}{600}; \\
Y.-S.~Wu, \Journal{\PRL}{73}{1994}{922}.

\bibitem{BCR}
P.~Bouwknegt, L.~Chim and D.~Ridout,
{\it Exclusion statistics in conformal field theory and the UCPF
for WZW models},
[{\tt hep-th/9903176}].

\bibitem{WZ}
X.-G.~Wen and A. Zee,  
\Journal{\PRB}{58}{1998}{15717}, [{\tt cond-mat/9711223}].

\bibitem{FNS}
E.~Fradkin, C.~Nayak and K.~Schoutens,
\Journal{\NPB}{546}{1999}{711}, [{\tt cond-mat/9811005}].

\end{thebibliography}
\end{document}